\documentclass[
reprint,
amsmath,amssymb,
aps,
prl
]{revtex4-2}

\usepackage{hyperref}
\hypersetup{colorlinks=true,
            linkcolor=black,
            citecolor=black,
            urlcolor=blue}
\usepackage{amssymb}
\usepackage{mathbbol}
\usepackage{courier}

\usepackage{graphicx}
\usepackage{dcolumn}
\usepackage{bm}

\begin{document}

\title{Surrogate model solver for impurity-induced superconducting subgap states}



\author{Virgil V. Baran$^{1,2,3,4}$}
\email[]{virgil.v.baran@unibuc.ro}
\author{Emil J. P. Frost$^{4}$}
\email[]{emiljpfrost@gmail.com}
\author{Jens Paaske$^{4}$}
\affiliation{$^{1}$Research Institute of the University of Bucharest (ICUB), 050107 Bucharest, Romania}
\affiliation{$^{2}$Faculty of Physics, University of Bucharest, 405 Atomi\c stilor, RO-077125, Bucharest-M\u agurele, Romania}
\affiliation{$^{3}$"Horia Hulubei" National Institute of Physics and Nuclear Engineering, 30 Reactorului, RO-077125, Bucharest-M\u agurele, Romania}
\affiliation{$^{4}$Center for Quantum Devices, Niels Bohr Institute, University of Copenhagen, 2100 Copenhagen, Denmark}

\begin{abstract}
A simple impurity solver is shown to capture impurity-induced superconducting subgap states in quantitative agreement with the numerical renormalization group and quantum Monte-Carlo simulations. The solver is based on the exact diagonalization of a single-impurity Anderson model with discretized superconducting reservoirs including only a small number of effective levels. Their energies and couplings to the impurity $d$-level are chosen so as to best reproduce the Matsubara frequency dependence of the hybridization function.
We provide a number of critical benchmarks and demonstrate the solvers efficiency in combination with the reduced basis method [Phys. Rev. B 107, 144503
(2023)] by calculating the phase diagram for an interacting three-terminal junction. 





\end{abstract}

\maketitle

Quantum dots (QD) tunnel-coupled to superconducting (SC) leads act as atomic Anderson impurities and induce Andreev bound states inside the superconducting gap. Depending on parameters, these subgap states range from Yu-Shiba-Rusinov (YSR) states~\cite{Yu1965, Shiba1968, Rusinov1969}, induced by odd occupied Coulomb blockaded QDs with a local magnetic moment, to a localized quasiparticle excitation above an induced gap on proximitized QDs with smaller charging energy~\cite{Bauer2007Nov, Meng2009Jun, Oguri2013, Kirsanskas2015}. These states are observed routinely either by scanning tunnelling spectroscopy (STS) near adatomic Anderson impurities on superconductor surfaces~\cite{Ruby2015, Ruby2016Oct, Choi2017May, Huang2020Dec, Villas2020Jun}, or in transport or microwave spectroscopy off semiconductor QDs contacted by superconductors~\cite{Grove-Rasmussen2009Apr, Maurand2012Feb, Lee2014Jan, Jellinggaard2016Aug, Delagrange2018May, Steffensen2022Apr, Bargerbos2022Jul, Fatemi2022Nov, Pita-Vidal2023May}. Understanding their detailed behavior is therefore of great importance for interpreting detailed STS subgap spectra to infer about the host superconductor, as well as for the design of superconducting qubits and other complex gateable superconductor-semiconductor hybrid devices relying on engineered subgap states~\cite{Mishra2021Dec, Pavesic2022Feb, Pita-Vidal2023May, Dvir2023Feb}.

The bound subgap states induced by an Anderson impurity in a superconductor can be calculated within a number of different approximate methods, which correspond reasonably well with the numerically exact results of numerical renormalization group (NRG)
or quantum Monte-Carlo (QMC) (cf. Refs.~\onlinecite{Martin-Rodero2011Oct,Meden2019Feb, Zalom2023Jul,Pokorny2023Apr} and references therein),
calculations in different regions of parameter space. These include the original YSR approach neglecting spin-flips~\cite{Yu1965, Shiba1968, Rusinov1969, Kirsanskas2015, Hermansen2022Feb}, the infinite-gap (atomic) limit \cite{Rozhkov2000Sep, Meng2009Jun, Hermansen2022Feb} and its recent generalizations \cite{Zonda2023Mar, Pokorny2023Apr}, as well as the zero bandwidth (ZBW) approximation \cite{Vecino2003Jul, Grove-Rasmussen2018Jun} of including only a single quasiparticle in the superconductor. 

\begin{figure}[t!]
\centering
\includegraphics[width=\columnwidth]{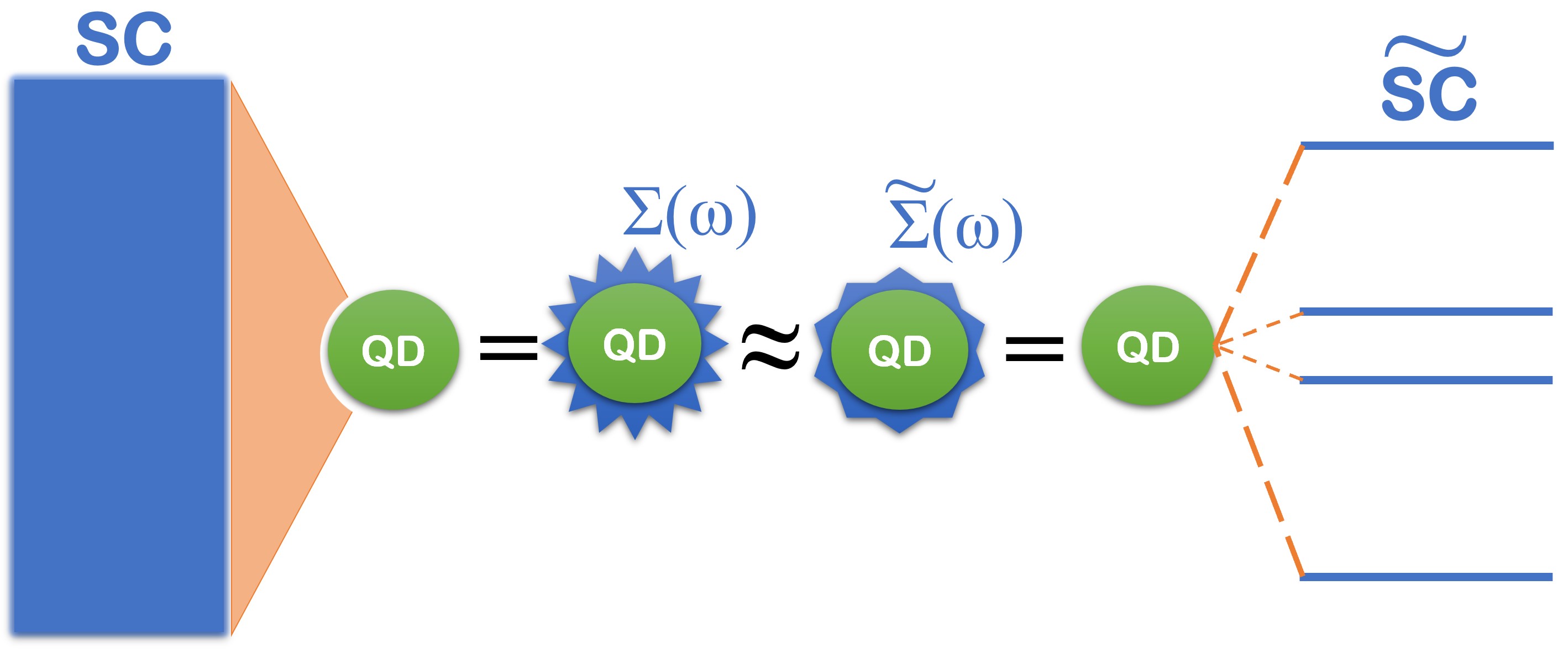}
     \caption{Schematic of the surrogate model. SC lead (blue continuum) coupled to a QD (green) by a given hybridization strength (orange), giving rise to the tunneling self-energy $\Sigma_d^{\text{T}}(\omega)$ of Eqs.~(\ref{Sigma},\ref{gcont}) (blue spikes). This self-energy is approximated by the prescription of Eq.~\eqref{tildeg}, corresponding to a few-level effective BCS model (blue $\widetilde{\text{SC}}$), depicted here with $\tilde{L}=4$ particle-hole symmetric levels coupled to the QD by different pairs of tunneling amplitudes (dashed orange lines).}              
\label{fig:schematic}
\end{figure}

In contrast to the normal state, the finite BCS gap in the superconducting quasiparticle excitation spectrum, $\Delta$, cuts off all logarithmic singularities and prevents an actual Kondo problem~\cite{Kondo1964Jul}. Unless its ratio to the Kondo temperature, $\Delta/T_{K}$, is very small, the superconducting gap therefore saves a lot of calculational effort, leaving a simpler non-perturbative problem of solving for bound states inside the gap. In more technical terms, the finite gap ensures that the normal component of the local BCS Nambu Green function, defining the tunneling self-energy $\Sigma_d^{\text{T}}(\omega_n)$, vanishes linearly with the Matsubara frequency below the gap. This has the convenient consequence that the same Green function can readily be obtained within a surrogate BCS model with a few discrete levels coupled to the $d$-level (cf. Fig.~\ref{fig:schematic}). Here we utilize this simplification to demonstrate that exact diagonalization (ED) of a low-dimensional surrogate BCS model coupled to the $d$-level captures the numerically exact results obtained by NRG and QMC.

In the context of Dynamical Mean Field Theory~\cite{Georges1996Jan}, ED has already been employed as an efficient impurity solver producing static thermodynamic quantities in very good agreement with QMC~\cite{Caffarel1994Mar, Koch2008Sep, Liebsch2011Dec, Mejuto-Zaera2020Jan}. Within any finite-size approximation, however, one loses the ability to properly describe the spectral function of the continuum, which is reduced to a collection of $\delta$-peaks (cf. e.g. Fig.~18 of Ref.~\cite{Georges1996Jan}). This limitation may be overcome by using an ensemble of discrete models within the so-called Distributional Exact Diagonalization approach~\cite{Granath2012Sep,Granath2014Dec, Motahari2016Dec}. Whereas this could be relevant for a faithful description of the above-gap continuum, this work will focus exclusively on the discrete subgap states and corresponding equilibrium expectation values of various observables.

\emph{Model}. We consider the superconducting Anderson impurity model Hamiltonian for a quantum dot (QD) coupled to a single superconducting lead (SC),
\begin{align}\label{ham0}
H&=H_{\text{QD}}+H_{\text{SC}}+H_{\text{T}},\\
     H_{\text{QD}}&=\epsilon_d \sum_{\sigma=\uparrow \downarrow} d^\dagger_\sigma d_\sigma+U d^\dagger_{\uparrow} d_{\uparrow} d^\dagger_{\downarrow}d_{\downarrow},\nonumber\\
     H_{\text{SC}}&=\sum_{\mathbf{k} \sigma} \xi_{ \mathbf{k}} c_{\mathbf{k} \sigma}^{\dagger} c_{ \mathbf{k} \sigma}- \sum_{\mathbf{k}}(\Delta e^{i\varphi}\,c_{\mathbf{k} \uparrow}^{\dagger} c_{-\mathbf{k} \downarrow}^{\dagger}+\text{h.c.}),\nonumber\\
         H_{\text{T}}&=\sum_{\mathbf{k} \sigma}(t c_{ \mathbf{k} \sigma}^{\dagger} d_{ \sigma}+\text {h.c.}).\nonumber
\end{align}
Here, $d^\dagger_\sigma$ creates an electron with spin $\sigma$ and energy $\epsilon_d$ on the QD 
with a repulsive on-site Coulomb interaction $U$. Similarly, $c^\dagger_{\mathbf{k}\sigma}$ creates an electron with spin $\sigma$, momentum $\mathbf{k}$ and energy $\xi_{\mathbf{k}}$ in the SC lead with order parameter $\Delta e^{i\varphi}$, where $\Delta$ and $\varphi$ are real numbers. The SC-QD tunnel coupling is described by $ H_{\text{T}}$, whose tunneling amplitudes, $t$, are taken to be momentum independent. We replace the momentum summation by an energy integral, assuming a constant density of states, $\nu_F=1/(2D)$, in a band of half-width $D$ around the Fermi surface. The discussion below may be trivially generalized beyond these simplifying assumptions. 

With the superconducting correlations being treated at the BCS mean-field level, the lead degrees of freedom are readily integrated out to give rise to the following Nambu tunneling self-energy (hybridization function)~\cite{Bauer2007Nov, Meng2009Jun}: 
\begin{align}\label{Sigma}
    \Sigma_d^{\text{T}}(\omega_n)=-\Gamma
    \begin{pmatrix}
            i\omega_n & \Delta e^{i\varphi}\\
            \Delta e^{-i\varphi} & i\omega_n 
     \end{pmatrix} g(\omega_n),
\end{align}
with Matsubara frequencies $\omega_n=(2n+1)\pi k_{\text{B}}T$ at temperature $T$, tunneling rate $\Gamma=\pi\nu_{F}|t|^{2}$, and the $g$-function defined as
\begin{align}\label{gcont}
g(\omega)&\equiv \frac{1}{\pi}\int_{-D}^D\text{d}\xi\,\frac{1}{\xi^2+\Delta^2+\omega^2}=\frac{2}{\pi}\frac{ \arctan\left(\frac{D}{\sqrt{\Delta^2+\omega^2}}\right)}{\sqrt{\Delta^2+\omega^2}}~,
\end{align}
which will be our main interest for the discretization procedure outlined below.

 \begin{figure}[ht!]
\centering
\includegraphics[width=\columnwidth]{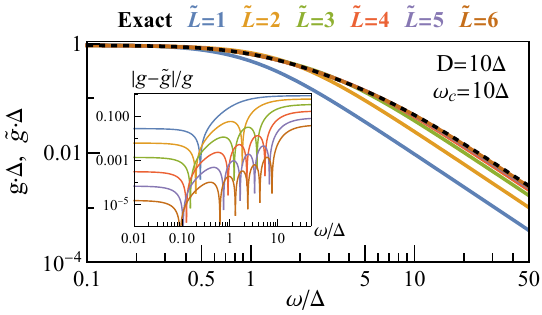}
\caption{The exact $g$-function (\ref{gcont}) (dashed black curve) and its best-fit approximations $\tilde g(\omega)$ (\ref{tildeg}) with $\tilde L\leq 6$. Inset: relative errors for the same parameters.} 
\label{fig2}
\end{figure}

\emph{Surrogate model}. We are interested in constructing the  simplest discrete effective bath that best reproduces the subgap states of the full model. For this purpose, we note that each SC level with energy $\xi$ contributes a factor of $(\xi^2+\Delta^2+\omega^2)^{-1}$ to the $g$-function (\ref{gcont}). Therefore, we seek to approximate the latter by combining only a small number of such factors,
\begin{equation}
\label{tildeg}
\begin{aligned}
    &\tilde{g}_{\text{even}}(\omega)\equiv2\sum_{\ell=1}^K  \frac{\gamma_\ell}{\tilde\xi_\ell^2+\Delta^2+\omega^2}~, ~\tilde L =2K~,\\
    &\tilde{g}_{\text{odd}}(\omega)\equiv\frac{\gamma_0}{\Delta^2+\omega^2}+\tilde{g}_{\text{even}}(\omega)~,~\tilde L =2K+1~ ,\\
    \end{aligned}
\end{equation}
where $K$ denotes the number of pairs of effective levels. 
A $\tilde g$-function of this form may be obtained by integrating out an effective superconducting bath with the same gap, $\Delta$, as the original one and whose $\tilde L$ discrete levels with energies $\pm|\tilde\xi_\ell|$ are coupled to the dot via a tunneling matrix elements $\tilde t_\ell=\sqrt{\gamma_\ell \Gamma}$. Note that each odd-$\tilde L$ model involves one extra level at zero energy, $\tilde \xi_0=0$. The effective bath is thus defined by parameters $\{\gamma_\ell, \tilde \xi_\ell\}$, which may be determined by one of the methods below.

The bath discretization strategies  developed so far in the literature are classified~\cite{deVega2015Oct} as direct discretization (standard in the context of NRG), orthogonal polynomial representation~\cite{Burkey1984Apr, Boehnke2011Aug}, and numerical optimization~\cite{Caffarel1994Mar, Gramsch2013Dec, Dorda2014Apr}. In the present work we opt for the latter approach, in which the parameters $\{\gamma_\ell, \tilde \xi_\ell\}$ are determined by minimizing the cost function $ \chi^2=\sum_j |g(\omega_j) - \tilde g(\omega_j)|^2$, which is evaluated on a non-uniformly spaced grid of frequencies $\omega_j$. To ensure a good fit of $g(\omega)$ at subgap frequencies, we use a grid of 1000 points logarithmically spaced in the interval $\omega\in [10^{-3}\Delta,\omega_c]$, with the cutoff frequency chosen to be of the order of the largest energy scale of the problem at hand, $\omega_c\sim\text{max}(\Delta, \Gamma,U)$. We found that a cutoff frequency $\omega_c=10\Delta$ was appropriate in all cases analyzed below (with $D=10\Delta$). Additional results for $D=10^2$-$10^5\Delta$ are discussed in the Supplemental Material (SM).

The exact  $g$-function (\ref{gcont}) is shown in Fig.~\ref{fig2}, together with the best-fit approximations $\tilde g(\omega)$ for $\tilde L \leq 6$, and their corresponding relative errors.
We notice that by increasing $\tilde L$ the errors are rapidly and systematically reduced by several orders of magnitude across (and beyond) the fitting range. 

Once a good fit has been found, the parameters $\{\gamma_\ell, \tilde \xi_\ell\}$ define the surrogate model Hamiltonian as a discretized version of Eq.~\eqref{ham0}, obtained by replacing the continuous momentum ${\bf k}$ by the discrete index $\ell$, with $\xi_{\bf k}\rightarrow\tilde\xi_\ell$ and $t\rightarrow\tilde t_\ell$. Extending the model to encompass multi-terminal systems with $N$ different SC leads, this amounts to
\begin{align}\label{hamtilde}
\widetilde{H}&=H_{\text{QD}}+\sum_{\alpha=1}^N \left(H_{\widetilde{\text{SC}},\alpha}+H_{\widetilde{\text{T}},\alpha}\right),\\
     H_{\widetilde{\text{SC}},\alpha}&=\sum_{\alpha\ell \sigma} \tilde\xi_{ \ell} c_{\ell\alpha \sigma}^{\dagger} c_{\ell\alpha \sigma}- \sum_{\ell=1}^{\tilde L}(\Delta_\alpha e^{i\varphi_\alpha}\,c_{\ell\alpha\uparrow}^{\dagger} c_{\ell\alpha \downarrow}^{\dagger}+\text{h.c.}),\nonumber\\
         H_{\widetilde{\text{T}},\alpha}&=\sum_{\ell=1}^{\tilde L}\sum_{ \sigma=\uparrow \downarrow}\sqrt{\gamma_\ell \Gamma_\alpha}( c_{ \ell\alpha \sigma}^{\dagger} d_{ \sigma}+\text {h.c.}),\nonumber
\end{align}
where  $c^\dagger_{a\alpha\sigma}$ creates an electron with spin $\sigma$ and energy $\tilde\xi_a$ in the SC lead $\alpha$ with  superconducting order parameter $\Delta_\alpha e^{i\varphi_\alpha}$. For simplicity, we restrict our attention to leads with identical gaps ($\Delta_\alpha=\Delta$), coupled with equal strength to the QD ($\Gamma_\alpha=\Gamma/N$). We solve for the low-lying eigenstates of the finite-sized effective models using exact diagonalization (ED) for $N=1,2$, or the density matrix renormalization group in the matrix-product-state formulation~\cite{White1992Nov, Schollwock2011Jan} for $N=2,3$. The latter is still an efficient solver for the (quasi-)1D systems considered here, and is straightforward to implement with the ITensor library~\cite{itensor,itensor-r0.3}. The methodology described in this section constitutes the proposed  surrogate model solver (SMS) for superconducting impurity problems. Our numerical codes are available online~\cite{github_ed, github} and may be run on a standard laptop or desktop computer.

\emph{Results}. First, we consider an S-D-S junction ($N=2$). At zero phase bias, this constitutes a single-channel problem where only the even ({\it gerade}) combination of quasiparticles from the two leads couple to the QD. The $\Gamma$ dependence of the excitation spectrum is shown in Fig.~\ref{fig3} for a moderately large value of the Coulomb interaction, at the particle-hole (ph) symmetric point. While the position of the singlet-doublet quantum phase transition is rather accurately captured even by the crudest $\tilde L=1$ (ZBW) approximation, deviations from the NRG data appear at higher energies in both the weak, and strong coupling regimes.

\begin{figure}[ht!]
\includegraphics[width=\columnwidth]{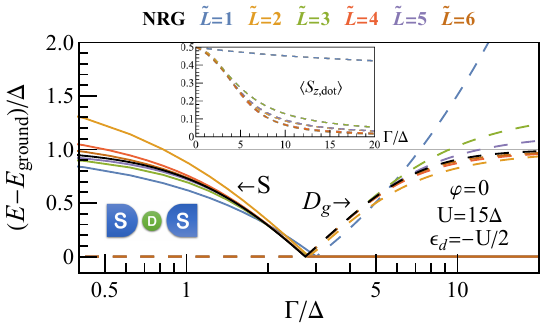}
\caption{SMS versus NRG \cite{peter_NRG} evolution of the subgap spectrum for an S-D-S junction with increasing coupling $\Gamma$ (log-scale). Continuous lines indicate the lowest singlet state $S$, dashed lines indicate the \emph{gerade} doublet $D_g$. Inset: average spin $S_z$ on the QD (in the $D_g$ state) versus increasing $\Gamma$.}
\label{fig3}
\end{figure}

The odd-$\tilde{L}$ surrogates always capture the correct singlet excitation energy, $E_{S}-E_{D}=\Delta$ at $\Gamma=0$, as they contain the $\tilde\xi_0=0$ level which accommodates a screening quasiparticle with energy $E_{qp}=\Delta$. For even $\tilde{L}$, $E_{qp}>\Delta$ since all levels have $|\tilde \xi_\ell|>0$, which leads to an overestimation of the singlet excitation energy. The situation at strong coupling is reversed, with even-$\tilde L$ surrogates performing well against the NRG data and the odd-$\tilde L$ ones overestimating the doublet excitation energy. This correlates with the inability of the odd-$\tilde L$ surrogates to properly screen the QD spin in the excited doublet ($D_{g}$), as illlustrated by the inset in Fig.~\ref{fig3}. Nevertheless, for large enough $\tilde L$, we obtain good convergence towards an excited doublet with a largely screened QD spin at strong coupling, in agreement with the findings of Ref.~\cite{Moca2021Oct, Pavesic2023Apr}.

\begin{figure}[t!]
\includegraphics[width=\columnwidth]{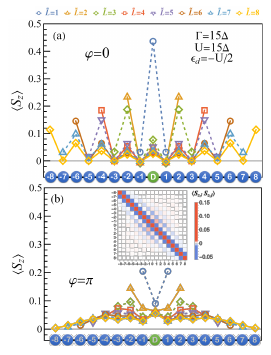}
\caption{Average spin $\langle S_z\rangle$ in the \emph{gerade} doublet $D_g$ state, across the chains representing the S-D-S junction at $\varphi=0$ (a) and $\varphi=\pi$ (b), for various numbers of effective levels $\tilde L\leq 8$ per lead, obtained for  $D=\omega_c=10\Delta$. The other common parameters are $\Gamma=15\Delta$, $U=15\Delta$, $\epsilon_d=-U/2$. Inset in panel (b): spin-spin correlation matrix $\langle S_{z,i} S_{z,j}\rangle$, $-\tilde L \leq i,j \leq \tilde L$, across the S-D-S chain at $\varphi=\pi$, where (blue) red squares indicate (anti-)aligned spins. The spin-spin correlation matrix is qualitatively similar in both singlet and doublet states.}
\label{fig4}
\end{figure}

To gain more insight into the surrogate-model eigenstates, it is advantageous to employ the {\it chain} representation of our discretized SC leads where the QD only couples to the first site in each SC chain. This picture is unitarily equivalent with the {\it star} configuration used so far, but it offers a more geometrical perspective in the interpretation of the impurity screening process. Unlike the DMRG calculations presented in Ref.~\cite{Moca2021Oct}, however, the small surrogate model chain has no precise sense of spatial distance in the real superconductors. A detailed derivation of the star-chain mapping can be found in Appendix A of Ref. \cite{Bulla2005Jan}.

Whereas the subgap singlet state is found to have $\langle S_{z,n}\rangle=0$ for all sites $n$ on the chain, Fig.~\ref{fig4} reveals a rich spin structure of the gerade doublet ($D_g$) state. The inset of Fig.~\ref{fig4}b shows pronounced antiferromagnetic correlations between nearest-neighbor sites of the S-D-S chain, typical to both singlet and doublet states. Qualitatively, the doublet spin structure may therefore be pictured as a delocalized $S_z=+1/2$ quasiparticle moving on top of a singlet background of antiferromagnetically correlated spins. For the one-channel problem ($\varphi=0$) in Fig.~\ref{fig4}a, the extra $S_z=+1/2$ spin is preferentially distributed on the sites ferromagnetically correlated with the dot (i.e. its even-order neighbors), with a remarkable tendency of localization towards the chain boundaries (away from the dot). In odd-$\tilde L$ chains, the boundary sites are antiferromagnetically correlated with the dot and can only accommodate a minute fraction of the total $S_z=+1/2$ spin. The latter must then be redistributed across the entire chain (predominantly on the dot and even SC sites), thus explaining the  poorly screened QD spin observed in the inset of Fig.~\ref{fig3}. The S-D-S junction at $\varphi=\pi$ (Fig.~\ref{fig4}b) features a qualitatively different spin structure, with minor differences between the spin distribution on neighboring sites; here even the $\tilde L=1$ (ZBW) approximation is qualitatively correct. Instead, the two-channel nature of the problem enables the formation of an extended spin-1/2 cloud around the impurity with negligible spin localized at the chain boundaries, visible in Fig.~\ref{fig4}b for $\tilde L\gtrsim 5$.

\begin{figure}[ht!]
\centering
\includegraphics[width=\columnwidth]{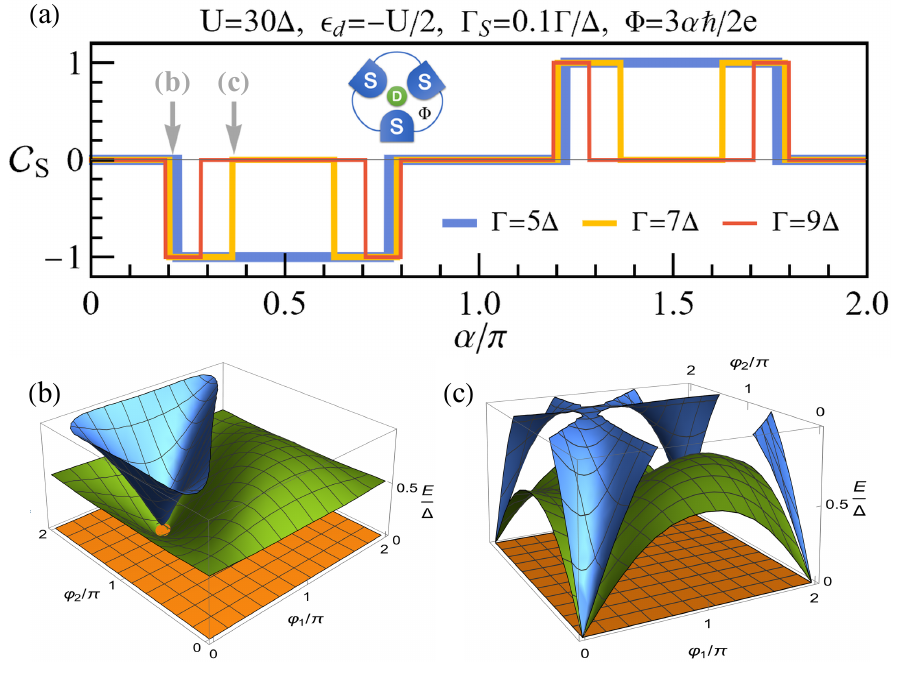}
\caption{Chern number and subgap spectrum in a three-terminal junction. (a) Singlet Chern number $\mathcal{C}_S$ as a function of the enclosed magnetic flux $\alpha$, for various QD-SC coupling strengths $\Gamma$. (b-c) Subgap energy spectrum near two singlet Weyl nodes, at $\alpha\approx 0.21\pi$(b) and $\alpha\approx 0.37\pi$ (c), obtained for $\Gamma=7\Delta$ with the $\tilde{L}=2$ surrogate. Orange indicates the lowest (reference) singlet, blue the first excited singlet, green the lowest doublet. For the efficient scan of the $\varphi_1-\varphi_2$ first Brillouin zone we employed the reduced-basis method, see Refs. \cite{Baran2023Apr, Herbst2022Apr, Brehmer2023Apr} and SM.}
\label{fig5}
\end{figure}

Having benchmarked the SMS, we shall now illustrate its capabilities towards solving a more complex topologically-nontrivial multi-terminal problem. Such devices are predicted to host Weyl points in a space of synthetic dimensions defined by their superconducting phases $\varphi_j$ (cf. Ref.~\cite{Teshler2023Apr} and references therein). Whereas earlier theoretical predictions have been obtained exclusively in the limit of infinite superconducting gaps~\cite{Klees2020May,Klees2021Jan,Teshler2023Apr}, the SMS allows us to explore for the first time the experimentally relevant case of finite gap and charging energy.

Here we consider the three-terminal setup of Ref.~\cite{Klees2020May} sketched in the inset of Fig.~\ref{fig5}a. For it to be topologically nontrivial, it is necessary to couple the SC terminals directly (we denote the corresponding tunneling rate by $\Gamma_S$), and to enclose a magnetic flux $\Phi\equiv3\alpha\hbar/2e$. The explicit Hamiltonian may be found in the Supplemental Material. A topological phase transition is signaled by a change in the Chern number, defined as the flux $\mathcal{C}(\alpha)\equiv(2\pi)^{-1}\int_{0}^{2\pi}\int_{0}^{2\pi} \text{d}\varphi_1 \text{d}\varphi_2\, (\partial_{\varphi_1} A_2-\partial_{\varphi_2} A_1)$ associated to the Berry connection $A_j=i\langle \psi| \partial_{\varphi_j}|\psi\rangle$ for a given (subgap) state $\psi$ (with $\varphi_3=0$).

The Chern number for the lowest singlet state $\mathcal{C}_S$ (computed numerically by the method of Ref. \cite{Fukui2005Jun}) is displayed in Fig.~\ref{fig5}a. Excellent convergence for this robust topological quantity was achieved already for $\tilde{L}=2$. The results indicate that the first significant change in the system's topology appears around $\Gamma=5\Delta$, when also tuning $\Gamma_S=0.1\Gamma/\Delta$. Here, new pairs of singlet Weyl nodes are found to emerge at $\varphi_{1,2}=0$, around $\alpha=(2k+1)\pi/2$, $k\in \mathbb{Z}$. They gradually and asymptotically migrate in $\alpha$, with increasing coupling, towards the previously known Weyl nodes (present also in the large-gap limit), shrinking the topologically nontrivial regions in the process. The presence of local Coulomb interactions causes only small quantitative changes in the above topological phase diagram (cf. SM for more details).

\emph{Conclusions}. This work shows that it is possible to capture efficiently the physical properties of the impurity-induced superconducting subgap states with a model involving a very small number of effective levels, chosen to reproduce the Matsubara frequency dependence of the superconducting hybridization function. The subgap spectrum and all related observables converge rapidly and systematically to NRG and QMC results with increasing number of levels. Various benchmarks against the latter~\cite{Zonda2016Jan,Karrasch2008Jan} are provided in the Supplemental Material, together with an additional benchmark against NRG of the Josephson current for an S-D-D-S junction~\cite{EstradaSaldana2018Dec,Zonda2023Mar}.  

The SMS provides easy access to the hybridization structure of the subgap states, offering new insights into the doublet spin distribution and the related impurity screening process, by going beyond the limitations of crudest ZBW and (generalized) atomic limit approaches without sacrificing the computational efficiency. Furthermore, the fast convergence of the SMS (cf. SM) enabled for the first time the solution of a fully interacting multi-terminal problem, difficult to approach by means of either NRG and QMC. By its simplicity and flexibility, the SMS may prove instrumental in exploring the fully interacting many-body physics of new complex hybrid devices, which are actively being pursued for quantum information processing.

\begin{acknowledgments}
\emph{Acknowledgments}. 
We thank R. Aguado, W. Belzig, L. Pave\v si\'c, G. Steffensen, P. Zalom and R. \v Zitko  for useful discussions and suggestions. This work was supported by grants of the Romanian Ministry of Education and Research, in part by Project No. PN-III-P4-ID-PCE-2020-1142, within PNCDI III, and in part by Project No. 760122/31.07.2023 within PNRR-III-C9-2022-I9. The Center for Quantum Devices is funded by the Danish National Research Foundation.
\end{acknowledgments}
\bibliography{apssamp}

\clearpage

\setcounter{figure}{0}
\renewcommand{\figurename}{FIG.}
\renewcommand{\thefigure}{S\arabic{figure}}

\renewcommand{\theequation}{S.\arabic{equation}}

\begin{widetext}
\begin{center}
    {\large{Supplemental Material for ``Surrogate model solver for impurity-induced superconducting sub-gap states"}}\\
\end{center}

We collect here additional supporting details regarding various aspects touched upon in the main text, and also a number of plots obtained for a larger bandwidth than that considered in the main text in Figs. S3-S8, the latter for an S-D-D-S junction.  We benchmark the surrogate models against QMC in Figs.~\ref{figS6},~\ref{figS7}. Their results at finite temperature $T$ have been obtained by combining the subgap state averages for the relevant observables with the appropriate $T$-dependent Boltzmann factors. Additional results for three-terminal junctions are provided in Figs. S11-S15.

\vspace{-0.5cm}

\subsection{Hybridization function fitting}

\vspace{-0.5cm}

\begin{table*}[hb!]\label{table1}

\begin{ruledtabular}
\begin{tabular}{ccccccccc}
 &$\tilde{L}=1$ &$\tilde{L}={2}$&$\tilde{L}={3}$ 
 &$\tilde{L}={4}$ & $\tilde{L}={5}$   & $\tilde{L}={6}$ & $\tilde{L}={7}$& $\tilde{L}={8}$\\
\hline
$\gamma_0$ & {0.9618} & & {0.5080} & & {0.3477} & & {0.2629}  \\
$\gamma_1$ & & {1.2642} & { 1.8454} & {0.5258} & {0.6371} & {0.3383} & {0.3728} & {0.2520}\\
$\gamma_2 $ & & & & {2.1140} & {2.1088} & {0.7476} &{ 0.8330} & {0.4182}\\
$\gamma_3$ & & & &   &  & {1.9703} &{1.7848} & {0.8896}\\
$\gamma_4 $ & & & &   &  &  &  &  {1.5940}\\
$\tilde\xi_1 $ & & {1.3099} & {2.7546} & { 0.7067} &{1.3742} & {0.4906} &{0.9369} & {0.3771}\\
$\tilde\xi_2 $ & & & & {4.1227} & {5.2473} & {2.0437} & {2.6982} & {1.3756}\\
$\tilde\xi_3$ & & & &   & &  {6.1272} & { 6.8092} & {3.3197} \\
$\tilde\xi_4 $ & & & &   & &   &   & {7.3402} \\
\end{tabular}
\end{ruledtabular}
\caption{Parameters $\{\gamma_\ell,\tilde\xi_\ell\}$ (in units of $\Delta$) corresponding to the best fit of the $g(\omega)$-function  with the discretized prescription $\tilde g(\omega)$ for $D=\omega_c=10\Delta$.}
\end{table*}

\begin{figure*}[ht!]

\includegraphics[width=0.8\textwidth]{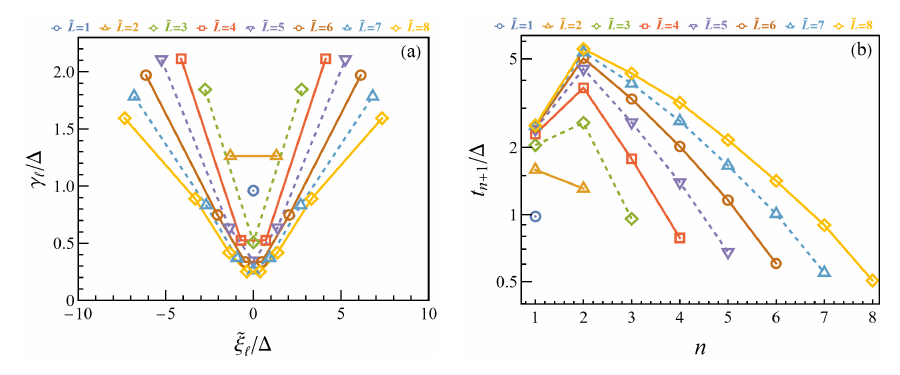}
\caption{ (a) Best-fit parameters $\{\gamma_\ell,\tilde\xi_\ell\}$ of Table I, with $\gamma_{-\ell}=\gamma_\ell$ and $\tilde\xi_{-\ell}=-\tilde\xi_\ell$. (b) Nearest neighbor hopping amplitudes $t_{n}$ (in log scale) across the equivalent QD-SC chain of eq. (\ref{chains}) for $\Gamma=\Delta$.}
\label{figS0}
\end{figure*}

 The fitting procedure described in the main text was found to produce the fastest convergence as long as $\Delta$ is not too small compared to $\Gamma$ and $U$, in which case also the high-frequency tail needs to be resolved well enough to capture the incipient Kondo correlations.  In each case, the optimal parameters $\{\gamma_\ell, \tilde \xi_\ell\}$ were obtained by minimizing $\chi^{2}$ using Mathematica's \texttt{NonlinearModelFit} routine~\cite{Mathematica}. Convergence for any observable should always be checked against increasing $\tilde L $ and $\omega_c$. 
 
Let us briefly comment on the general properties of the best-fit parameters $\{\gamma_\ell,\tilde\xi_\ell\}$. As a consequence of the low frequency range being favored by our fitting procedure, there is a very good agreement $\tilde g (\omega=0)\simeq g(\omega=0)$. This amounts, e.g. for even-$\tilde L$,
to $\sum_{\ell=1}^{\tilde L/2} \gamma_\ell/(\tilde\xi_\ell^2+\Delta^2)\simeq [\text{arctan}(D/\Delta)]/(\pi\Delta)\, $. While the lowest lying surrogate levels contribute the most to the left hand side of this expression, the higher lying levels ensure that the high-frequency tail of $g(\omega)$ is also well reproduced. For $\omega\gg D, \Delta,\tilde\xi_\ell$, both $\tilde g(\omega)$ and $g(\omega)$ decay as $1/\omega^2$, with the corresponding coefficients satisfying, e.g. for even-$\tilde L$, $\sum_{\ell=1}^{\tilde L/2} \gamma_\ell \approx D/\pi$. We may identify the effective bandwidth of the surrogate model as $\tilde D=\pi \sum_{\ell=1}^{\tilde L/2}\gamma_\ell$, to which the higher lying levels contribute the most. 

\subsection{Critical current of an S-D-S Junction}

 \begin{figure}[ht!]
\includegraphics[width=0.5\columnwidth]{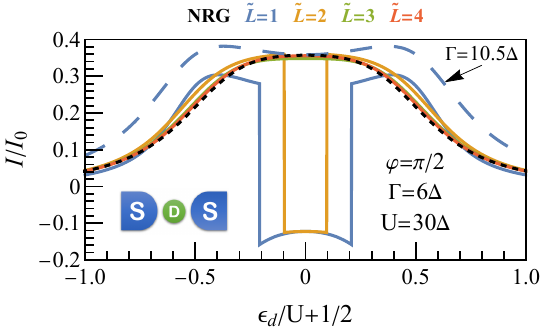}
\caption{Ground state Josephson current (in units of $I_0=e\Delta/\hbar$) versus $\epsilon_d$. Black curves indicate NRG data extracted from Ref.~\cite{rokdata}, while colored curves correspond to the surrogate models with $\tilde L\leq 4$ effective levels per lead, obtained for $D=\omega_c=10\Delta$.}
\label{fig3}
\end{figure}

At finite phase bias, the S-D-S junction supports a Josephson current, which may be calculated from the surrogate model ground state as $I=(2e/\hbar)\langle \partial_{\varphi}\widetilde{H}\rangle$. The result is plotted against $\epsilon_{d}$ in Fig.~\ref{fig3} at moderately strong coupling and strong Coulomb interaction. While the NRG data shows a smooth profile corresponding to a global singlet ground state, the $\tilde L=1,2$ surrogates show discontinuities associated with spurious singlet-doublet transitions. These are resolved for $\tilde L\geq 3$, leading to an excellent quantitative match with the NRG data. We note, that the qualitative failure  of the $\tilde L=1$ (ZBW) approximation in this regime cannot be remedied by a rescaling of $\Gamma$, which is often the least known parameter in an actual experiment. Even if the coupling is increased to produce a global singlet ground state, the current still shows two pronounced false maxima away from the ph-symmetric point (see the blue dashed line in Fig.~\ref{fig3}). Their presence is not uncommon in the NRG data for the singlet current at weaker couplings, however they are always hidden within the regions of doublet ground state. Also this example illustrates that the commonly employed ZBW approximation should be used with care when $\Delta$ is much smaller than $\Gamma$ and $U$.

\subsection{Surrogate Hamiltonian in the chain representation}

The equivalent chain representation of the surrogate Hamiltonian of eq. (5) in the main text is given by 
\begin{align} \label{chains}
     H_{\widetilde{\text{T}},\alpha} + H_{\widetilde{\text{SC}},\alpha} &= \sum_{ \sigma=\uparrow \downarrow} (t_{0,\alpha} f^\dagger_{1,\alpha\sigma}d_{\sigma}+\text{h.c.})+\sum_{n=1}^{\tilde L-1}\sum_{ \sigma=\uparrow \downarrow} (t_{n\alpha} f_{n+1,\alpha\sigma}^{\dagger} f_{n\alpha\sigma} +\text{h.c.}) - \sum_{n=1}^{\tilde L}(\Delta_\alpha e^{i\varphi_\alpha}\,f_{n\alpha\uparrow}^{\dagger} f_{n\alpha\downarrow}^{\dagger}+\text{h.c.})\,,
\end{align}
where  $f^\dagger_{n\alpha\sigma}$ creates a fermion with spin $\sigma$  on the $n$'th site of the SC chain $\alpha$ with nearest-neighbor hopping amplitudes $t_{n\alpha}$. For each lead, the {\it chain} $f$-operators are related to the {\it star} $c$-operators via a unitary transformation $f^\dagger_{n\alpha\sigma}=\sum_{\ell=1}^{\tilde L} U^{(\alpha)}_{n\ell} c^\dagger_{\ell\alpha\sigma}$, which can be found by a tridiagonalization procedure~\cite{Bulla2005Jan}. The nearest-neighbor hopping amplitudes in the chain-representation of an S-D junction are shown in Fig. \ref{figS0}b. For $n > 2$, the hopping amplitudes fall off exponentially away from the QD.

\vspace{-0.5cm}

\newpage

\subsection{Efficient one-channel surrogate models for multi-terminal single-impurity problems }

We expand here on the possibility of mapping a multi-channel single-impurity problem down to a single channel. Consider the dot tunneling self-energy in the presence of $N$ superconducting leads with order parameters $\Delta e^{i\varphi_n}$, each coupled to the QD with a hybridization strength $\Gamma_n$,
\begin{align}
    \Sigma_d^{\text{T}}(\omega)=-\sum_{n=1}^N\Gamma_n
    \begin{pmatrix}
            i\omega & \Delta e^{i\varphi_n}\\
            \Delta e^{-i\varphi_n} & i\omega
     \end{pmatrix} g(\omega)\equiv -\Gamma
    \begin{pmatrix}
            i\omega & \widetilde\Delta e^{i\varphi}\\
            \widetilde\Delta e^{-i\varphi} & i\omega
     \end{pmatrix} g(\omega)~,
\end{align}
where we introduced the total hybridization strength $\Gamma\equiv\sum_{n=1}^N\Gamma_n$ and the effective gap $\widetilde\Delta e^{i\varphi}\equiv\sum_{n=1}^N \Delta e^{i\varphi_n} \Gamma_n/\Gamma$, with $0\leq \widetilde\Delta \leq\Delta$. While the integral representation of $g(\omega)$ in the main text still involves the original gap $\Delta$, it may nevertheless be cast into a form involving $\widetilde\Delta$ by a simple change of variables that preserves the quasiparticle excitation energy $E_{\text{qp}}^2=\xi^2+\Delta^2\equiv\zeta^2+\widetilde\Delta^2$,
\begin{align}
     g(\omega)=\frac{1}{\pi}\int_{-D}^D\text{d}\xi\,\frac{1}{\xi^2+\Delta^2+\omega^2}=\frac{1}{\pi}\int_{\mathcal{I}_\zeta}\text{d}\zeta\,\frac{\zeta}{\xi(\zeta)}\,\frac{1}{\zeta^2+\widetilde\Delta^2+\omega^2}~,\quad  \raisebox{-65pt}{\includegraphics[width=0.25\textwidth]{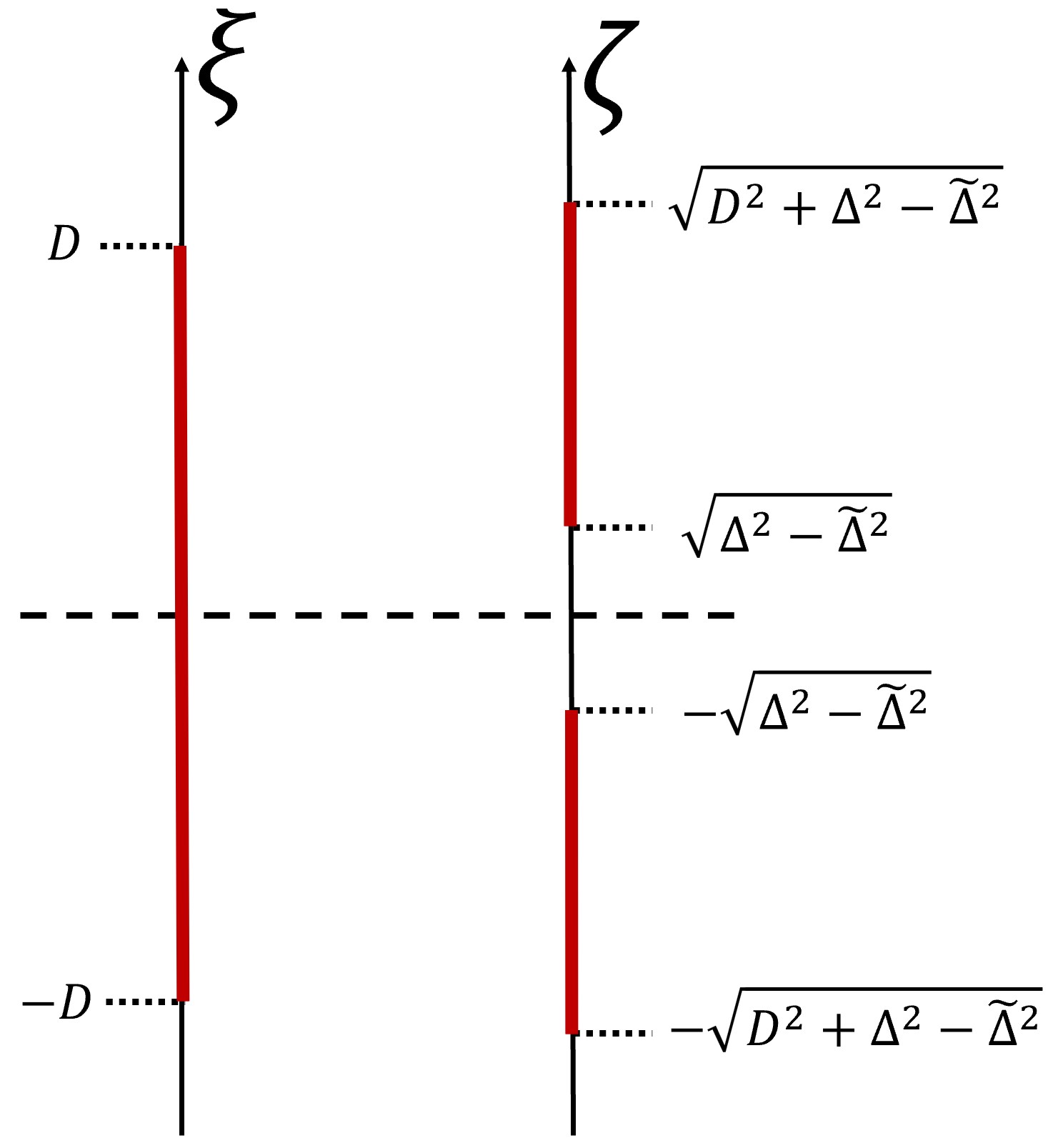}}
\end{align}
where $\xi(\zeta)=[\text{sgn}(\zeta)]\cdot [\zeta^2-(\Delta^2-\widetilde\Delta^2)]^{1/2}$. The equivalent integration domains $\mathcal{I}_\xi=[-D,D]$ and $\mathcal{I}_\zeta$ are highlighted in red in the diagram. The previous two expressions may be combined in 
\begin{align}
    \Sigma_d^{\text{T}}(\omega)=-\frac{\Gamma}{\pi}\int_{\mathcal{I}_\zeta}\text{d}\zeta\,\frac{\zeta}{\xi(\zeta)}\,
    \left[
    \begin{pmatrix}
            i\omega & \widetilde\Delta e^{i\varphi}\\
            \widetilde\Delta e^{-i\varphi} & i\omega
     \end{pmatrix} \frac{1}{\zeta^2+\widetilde\Delta^2+\omega^2}\right]=-\frac{\Gamma}{\pi}\int_{\mathcal{I}_\zeta}\text{d}\zeta\,\frac{\zeta}{\xi(\zeta)}\,
    \left[ -i\omega+
    \begin{pmatrix}
           \zeta & \widetilde\Delta e^{i\varphi}\\
            \widetilde\Delta e^{-i\varphi} & -\zeta
     \end{pmatrix} \right]^{-1}~,
\end{align}
which may be easily interpreted as a one-channel tunneling self-energy arising from integrating out an effective SC lead with order parameter $\widetilde\Delta e^{i\varphi}$, coupled to the impurity with an energy-dependent tunneling amplitude, $t(\zeta)=\zeta/\xi(\zeta)$. \\

To build the effective one-channel surrogate model, we simply rewrite our usual discretized $\tilde g(\omega)$ into a form displaying explicitly the effective gap $\widetilde\Delta$, e.g. for even-$\tilde L$ we have
\begin{equation}
\begin{aligned}
    &\tilde{g}(\omega)=2\sum_{\ell=1}^{\tilde L/2}  \frac{\gamma_\ell}{\tilde\xi_\ell^2+\Delta^2+\omega^2}=2\sum_{\ell=1}^{\tilde L/2}  \frac{\gamma_\ell}{\tilde\zeta_\ell^2+\widetilde\Delta^2+\omega^2}~,~\text{such that}~ E_{\text{qp},\ell}^2=\tilde\xi_\ell^2+\Delta^2\equiv\tilde\zeta_\ell^2+\widetilde\Delta^2~,
    \end{aligned}
\end{equation}
with effective level positions $\tilde\zeta_\ell=\pm\sqrt{\tilde\xi_\ell^2+\Delta^2-\widetilde\Delta^2}$. Note that, for $\widetilde \Delta=0$ the effective one-channel surrogate model actually involves only non-superconducting levels, but there is still no low-frequency divergence of the tunneling self-energy due to the constraint $|\tilde\zeta_\ell|\geq\Delta$ valid in this case. 

Explicitly, the effective one-channel surrogate Hamiltonian is 
\begin{align}
\widetilde{H}&=H_{\text{QD}}+\sum_{\ell=1}^{\tilde L}\sum_{ \sigma=\uparrow \downarrow}\tilde\zeta_{ \ell} c_{\ell \sigma}^{\dagger} c_{\ell \sigma}- \sum_{\ell=1}^{\tilde L}(\widetilde\Delta e^{i\varphi}\,c_{\ell\uparrow}^{\dagger} c_{\ell \downarrow}^{\dagger}+\text{h.c.})+\sum_{\ell=1}^{\tilde L}\sum_{ \sigma=\uparrow \downarrow}\sqrt{\gamma_\ell \Gamma}( c_{ \ell \sigma}^{\dagger} d_{ \sigma}+\text {h.c.}),
\end{align}
For the same number $\tilde L$ of levels per lead, the effective one-channel surrogates are computationally less expensive to solve than their multi-channel counterparts, and do not show any loss in  the accuracy of the results (see e.g. Fig. \ref{figS4} below).

\vfill
\newpage

\subsection{Additional numerical results}

\begin{figure*}[ht!]
\includegraphics[width=0.8\textwidth]{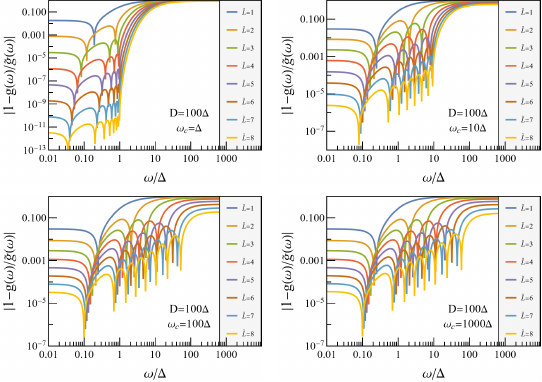}
\caption{Errors for the approximate $\tilde g(\omega)$-functions relative to the exact $g(\omega)$  with $\tilde L\leq 8$, for $D=100\Delta$ at various $\omega_c=1,10,100,1000\Delta$.}
\label{figS1}
\end{figure*}

\begin{figure*}[ht!]
\includegraphics[width=0.8\textwidth]{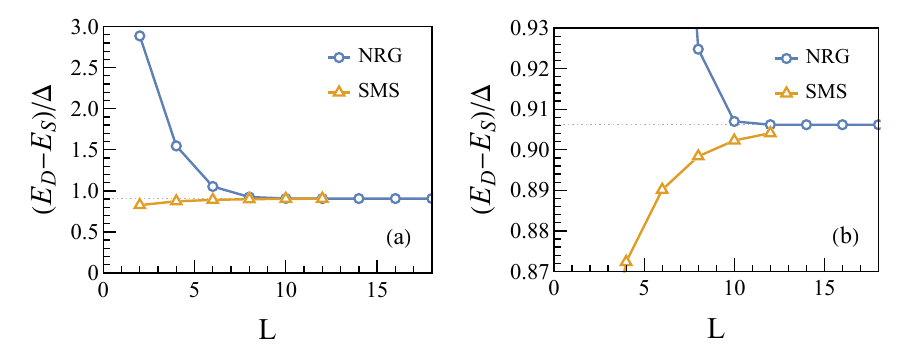}
\caption{Subgap state energy convergence for SMS and standard NRG for the S-D-S system at $\varphi=0$, $\Gamma=10\Delta$, $U=15\Delta$, $\epsilon_d=-U/2$, $D=10\Delta$ (same parameters as in Fig. 3 of the main text). We are grateful to P. Zalom for providing us with the results of the NRG Ljubljana implementation \cite{zitko_rok_2021_4841076} presented here. For $\varphi=0$, two-terminal superconducting single impurity Anderson model becomes effectively of one-channel nature. The corresponding one-channel NRG implementation has then been employed at $\Lambda=2$ with $500$ kept states (no z-averaging has been employed).}
\label{fig_p}
\end{figure*}

\end{widetext}

\clearpage
\begin{figure}
\includegraphics[width=\columnwidth]{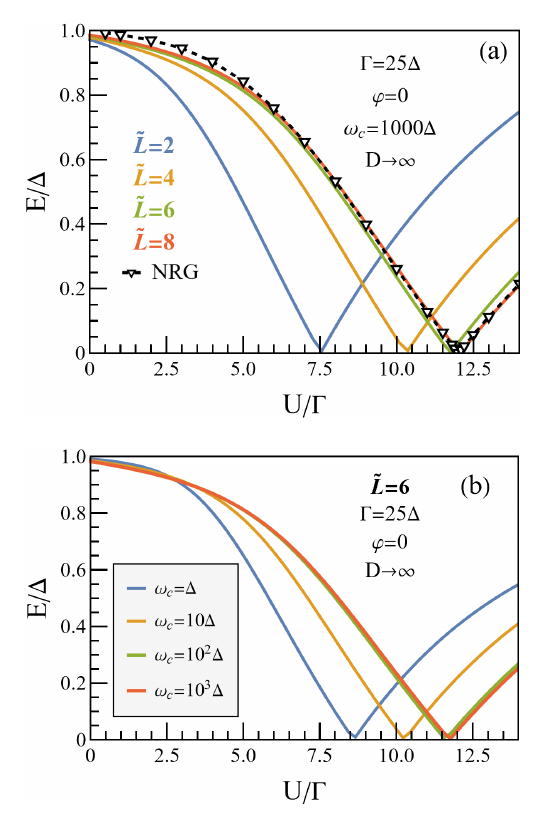}
\caption{S-D-S sub-gap state energy $E$ (in units of $\Delta$) vs the interaction strength $U$ (in units of $\Gamma$), at half-filling for different numbers of effective levels $\tilde{L}=2,4,6$ at $\Delta=0.04\Gamma$ and $\omega_c=100\Delta$ (a), and for different cutoff frequencies $\omega_c=10^k \Delta$, $k=0,1,2,3$ at $\tilde{L}=6$ and $\Delta=0.04\Gamma$ (b).  In each case, the $g(\omega)$-fit has been performed on 1000 logarithmically spaced points in the interval $\omega\in [10^{-3}\Delta,\omega_c]$, with infinite bandwidth $D\rightarrow\infty$. The NRG values have been read  graphically from Fig. 7(a) of Ref. \cite{Zonda2016Jan}.}
\label{figS2}
\end{figure}

\begin{figure}
\includegraphics[width=\columnwidth]{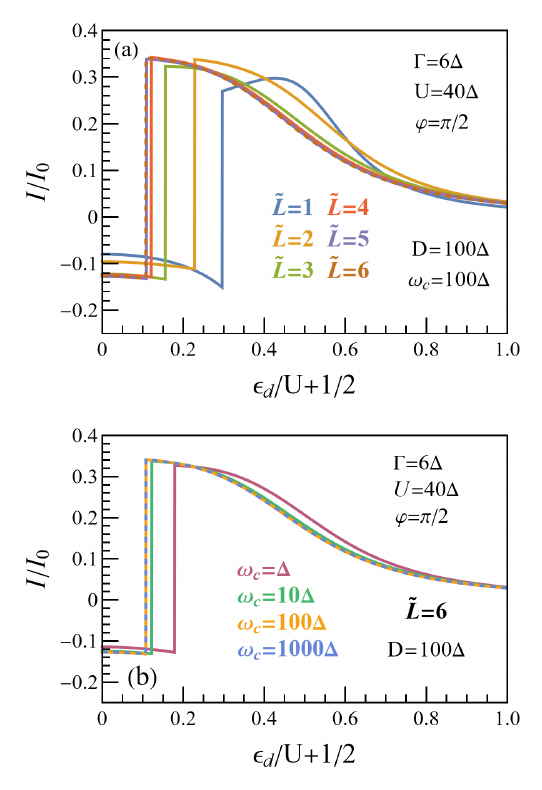}
\caption{(a) S-D-S Josephson current  $I$ (in units of $I_0=e\Delta/\hbar$) vs. the dot level position $\epsilon_d $, for different numbers of effective levels $\tilde{L}\leq 6$ at $U=40\Delta, \Gamma=6\Delta, \varphi =\pi/2$, $D=\omega_c=100\Delta$. (b) Convergence of the Josephson current with increasing cutoff frequency $\omega_c=1,10,100,1000\Delta$, for $\tilde L=6$ with the same parameter values as in (a).}
\label{figS3}
\end{figure}
 
\clearpage
    
\begin{figure}

\hspace{1.6cm}\includegraphics[width=0.15\textwidth]{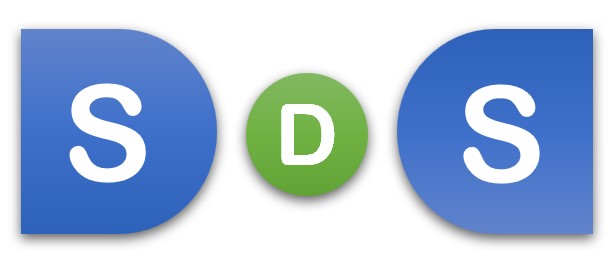}
\includegraphics[width=\columnwidth]{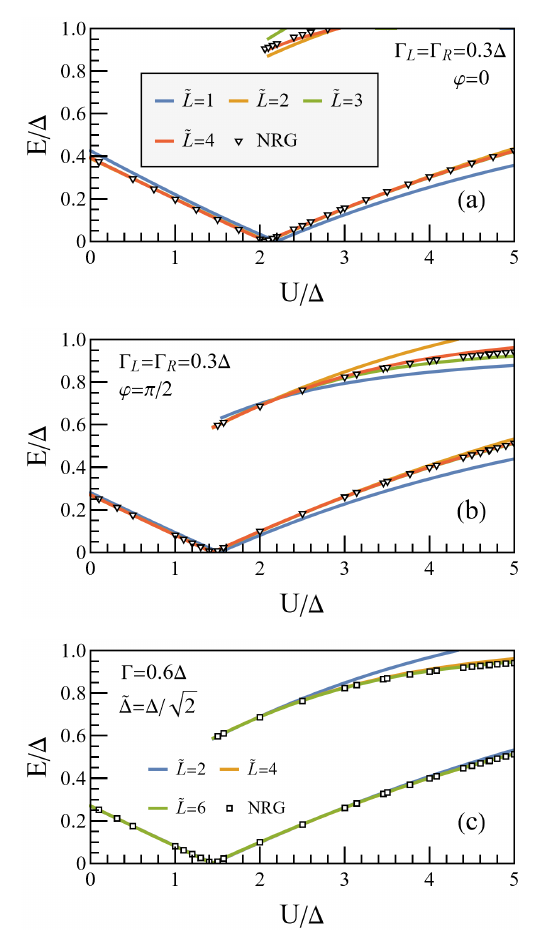}
\caption{S-D-S sub-gap state energies $E$ as functions of the interaction strength $U$, in units of $\Delta$, at half-filling with $\Gamma_L=\Gamma_R=0.3\Delta$, $\varphi=0$ (a) and $\varphi=\pi/2$ (b, same legend as in a), for various numbers of effective levels per lead $\tilde{L}\leq 4$.  The $g(\omega)$-fit has been performed on 1000 logarithmically spaced points in the interval $\omega\in [10^{-3}\Delta,\Delta]$, with bandwidth $D=100\Delta$. The NRG values have been read  graphically from Fig. 13 of Ref. \cite{Pokorny2023Apr}. Panel (c) shows the same case as in (b), but the results correspond to the effective one-channel surrogate models with effective gap $\widetilde\Delta=\Delta/\sqrt{2}$ (see discussion above).}
\label{figS4}
\end{figure}

\begin{figure}
\centering
\hspace{1.6cm}\includegraphics[width=0.15\textwidth]{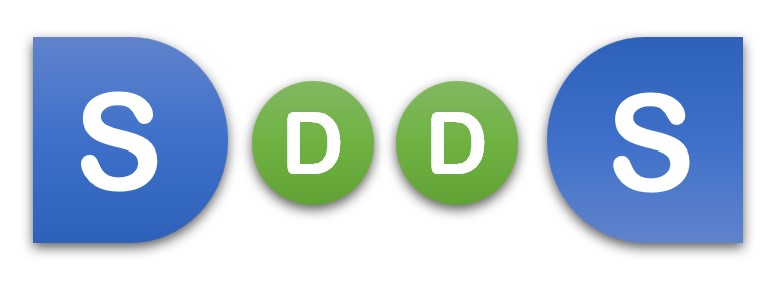}
\includegraphics[width=\columnwidth]{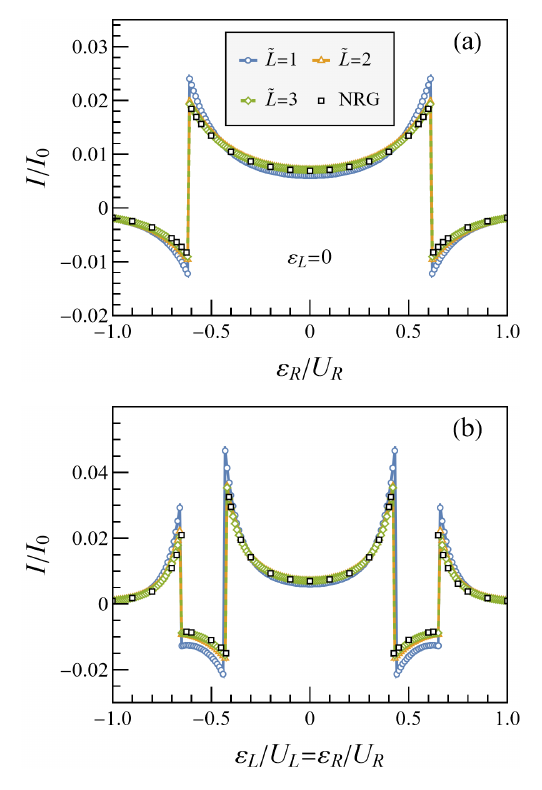}
\caption{Josephson current $I$ (in units of $I_0=e\Delta/\hbar$) for the S-D-D-S junction of Ref. \cite{Zonda2023Mar} vs. the shifted level positions $\varepsilon_{L,R}=\epsilon_{L,R}+U_{L,R}/2$, for $\varepsilon_L=0$ (a) and $\varepsilon_L/U_L=\varepsilon_R/U_R$ (b), at $U_L=7\Delta, U_R=6\Delta, t_d=1.2\Delta, \Gamma_L=\Gamma_R=0.3\Delta, \varphi =\pi/2$.  The $g(\omega)$-fit has been performed on 1000 logarithmically spaced points in the interval $\omega\in [10^{-3}\Delta,\Delta]$, with bandwidth $D=100\Delta$. The NRG values have been read  graphically from Fig. 13 of Ref.  \cite{Zonda2023Mar}.}
\label{figS5}
\end{figure}

\clearpage

\begin{widetext}

\begin{figure}

\centering
\includegraphics[width=0.075\textwidth]{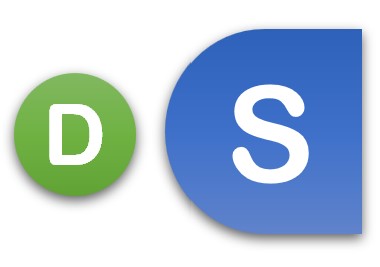}

\vspace{0.5cm}

\includegraphics[width=0.8\textwidth]{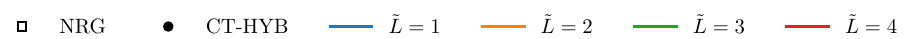}

\vspace{0.5cm}

\includegraphics[width=0.8\textwidth]{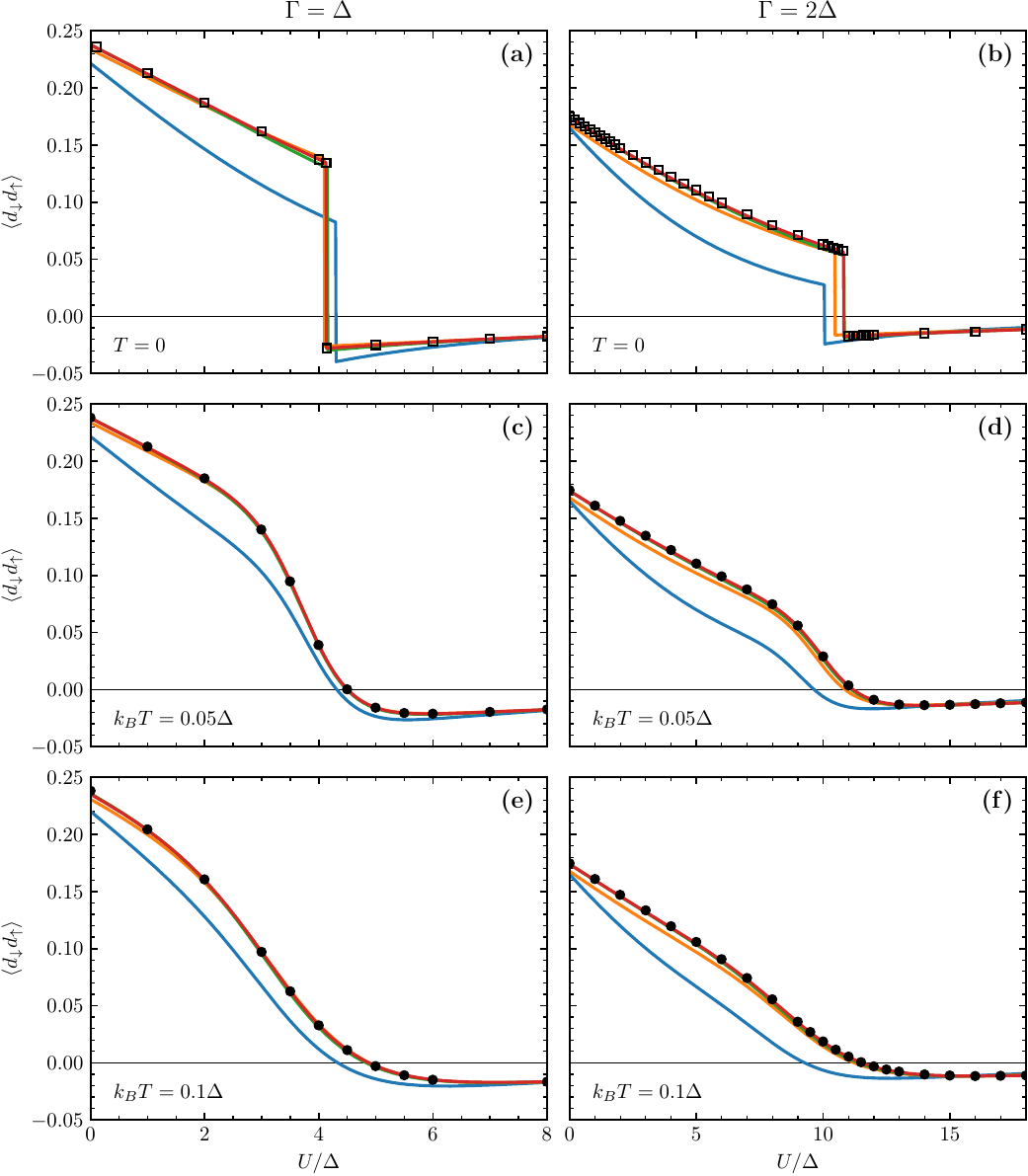}
\caption{D-S induced dot pairing $\langle d_\downarrow d_\uparrow\rangle$ as function of the interaction strength $U$ (in units of $\Delta$) for various numbers of effective levels $\tilde{L}=1,2,3, 4$, at half-filling with $\Gamma=\Delta$ (a, c, e) and $\Gamma = 2 \Delta$ (b, d, f) at different temperatures: $T=0$ (a, b), $k_BT=0.05\Delta$ (c, d), and $k_BT=0.1\Delta$ (e, f). The $g(\omega)$-fit has been performed on 1000 logarithmically spaced points in the interval $\omega / \Delta \in [10^{-3}, 10]$, with bandwidth $D=100\Delta$. The NRG and CT-HYB values have been read  graphically from Fig. 2(b, d) of Ref. \cite{Pokorny2023Apr}.}
\label{figS6}
\end{figure}

\clearpage

\begin{figure}
\centering
\includegraphics[width=0.125\textwidth]{sds.jpg}

\vspace{0.5cm}

\includegraphics[width=0.8\textwidth]{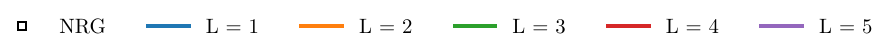}

\vspace{0.5cm}

\includegraphics[width=0.8\textwidth]{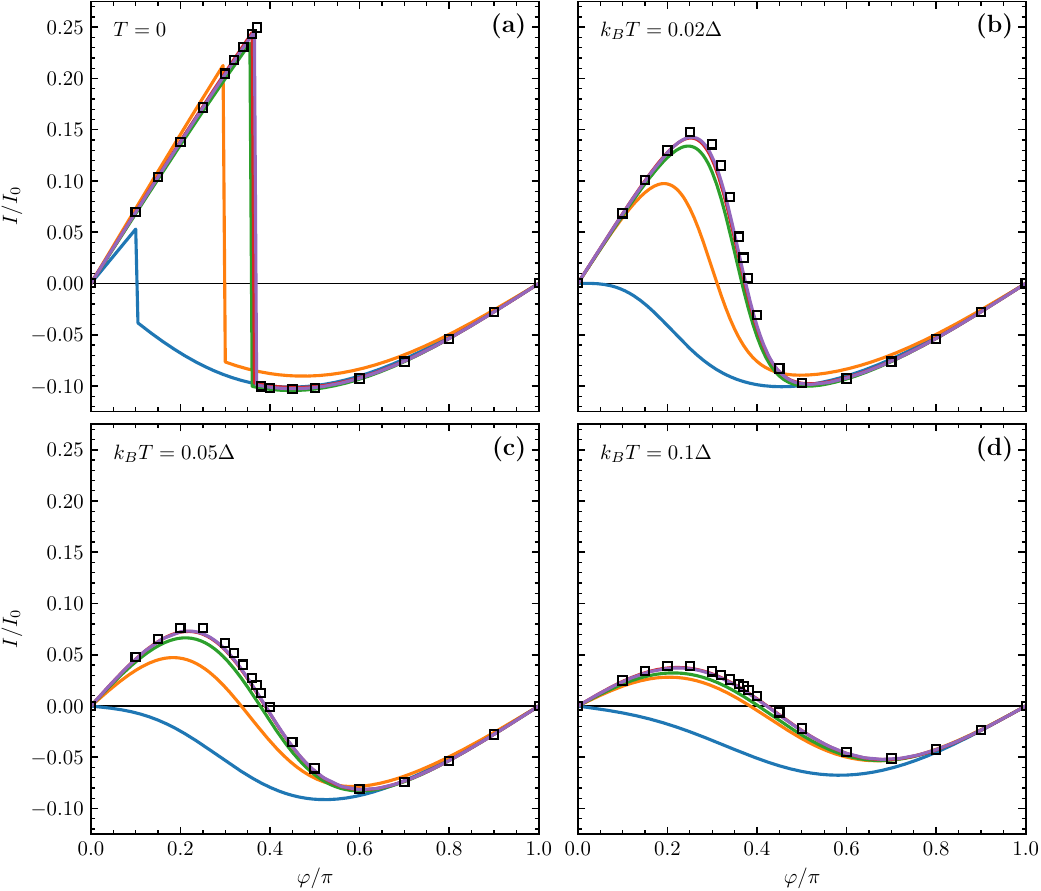}
\caption{S-D-S Josephson current (in units of $I_0 = e \Delta / \hbar$) as a function of phase bias $\varphi$ for various numbers of effective levels per lead $\tilde{L}=1,2, ..., 5$ at half-filling with $U = 5.2 \Gamma$ and $\Delta = 0.37\Gamma$ at different temperatures: $T=0$ (a), $k_BT=0.02\Delta$ (b), $k_BT=0.05\Delta$ (c), and $k_BT=0.1\Delta$ (d). The $g(\omega)$-fit has been performed on 1000 logarithmically spaced points in the interval $\omega / \Delta \in [10^{-3}, 10]$, with bandwidth $D=70.25\Delta$. The NRG values have been read graphically from Fig. 10 of Ref. \cite{Karrasch2008Jan}.}
\label{figS7}
\end{figure}

\clearpage

\subsection{Chimney diagrams and Reduced-basis Methodology }

\begin{figure}[ht!]
\centering
\includegraphics[width=0.6\columnwidth]{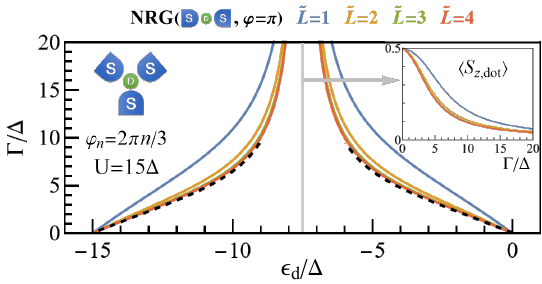}
\caption{``Chimney" stability  diagram of a phase-biased 3-terminal junction obtained by the RB-DB surrogates with $\tilde L\leq4$ levels per lead, for $D=\omega_c=10\Delta$.  The inner (outer) region displays a doublet (singlet) ground state. The NRG data corresponding to the latter has been extracted from Ref. \cite{rokdata} and is shown by the black dashed curves. Inset: average dot spin $S_z$ in the \emph{gerade} doublet $D_g$ ground state versus increasing coupling at  $\epsilon_d=-U/2$.}
\label{figS10}
\end{figure}

For the two-channel S-D-S $\varphi=\pi$ problem, the surrogate solver converges rapidly to the well-known {\it chimney} phase diagram~\cite{Rozhkov2000Sep, Tanaka2007May, Bargerbos2022Jul, Pavesic2023Apr} shown in Fig.~\ref{figS10}.  More specifically, in light of the main text discussion regarding the spin structure of the S-D-S junction at $\varphi=\pi$, it is not surprising to find good qualitative agreement for all surrogates regarding the singlet-doublet phase boundary, as well as the screening of the QD spin, with excellent quantitative agreement being reached already for $\tilde L=3$-$4$.

The results displayed in Fig.~\ref{figS10} are actually obtained for a more complex three-terminal junction, phase biased with $\varphi_n=2\pi n/3$, $n=1,2,3$, which exhibits the same total tunneling self-energy, $\Sigma_d^{\text{T}}(\omega_n)=-i\omega_n \Gamma g(\omega_n) I_{2\times 2}$, as the S-D-S junction with $\varphi=\pi$. We choose to solve this multi-terminal case by brute force to emphasize the surrogate model's applicability to more general multi-channel problems where an efficient single-channel picture may not always be available (see Refs. \cite{Zalom2021Jan, Zalom2021Jul} and discussion above for examples of multi- to one-channel mappings).

Instead, to address the increased computational complexity inherent to generic multi-channel problems, we propose the use of the reduced-basis method (RBM) \cite{Baran2023Apr, Herbst2022Apr, Brehmer2023Apr} to accelerate the scan of the relevant parameter space, e.g. the $\Gamma$-$\epsilon_d$ region shown in  Fig.~\ref{figS10}.  The basic RBM idea is to exploit the high linear dependence of a finite system's state vectors across any given region in parameter space. Concretely, the full many-body problem for a discretized-bath surrogate model is projected down to a reduced subspace spanned by its subgap state vectors computed by ED or DMRG for a small number of select parameter values (typically a few tens). As the RBM offers an interpolation of the state vector itself, any observable may be evaluated efficiently by its RB representation, e.g. the impurity spin shown in the inset of Fig.~\ref{figS10}. These reduced-basis surrogates allow for a fine scan across any region of the parameter space, several orders of magnitude faster than a brute force ED/DMRG scan on the same fine grid.

The reduced basis is constructed iteratively, at each stage computing a new subgap state vector (using DMRG) at the position in the parameter space of the maximum residual Res=$||H|\psi^{(\text{rb})}\rangle-E^{(\text{rb})}_{\text{subgap}}|\psi^{(\text{rb})}\rangle ||$, where the ``rb" indexed quantities are obtained by solving the Hamiltonian eigenvalue problem in the reduced subspace. The residual may be evaluated efficiently across the entire parameter region of interest using the RB already assembled, and acts as a local measure for the RB emulation error. In obtaining Fig.~\ref{figS10}, we required that the residual should not exceed the value Res$_{\text{max}}=0.01\Delta$ on a uniform discretization in 40$\times$40 grid points of the region $(\Gamma/\Delta,\epsilon_d/\Delta)\in [0,20]\times[-7.5,3.75]$. The left half of Fig.~\ref{figS10} was then constructed by ph symmetry. The largest reduced basis dimension obtained was a modest $N_{\text{rb}}=25$ (in each parity sector) for the largest $\tilde L=4$ surrogate considered.

\end{widetext}
\clearpage

\begin{figure}
\includegraphics[width=\columnwidth]{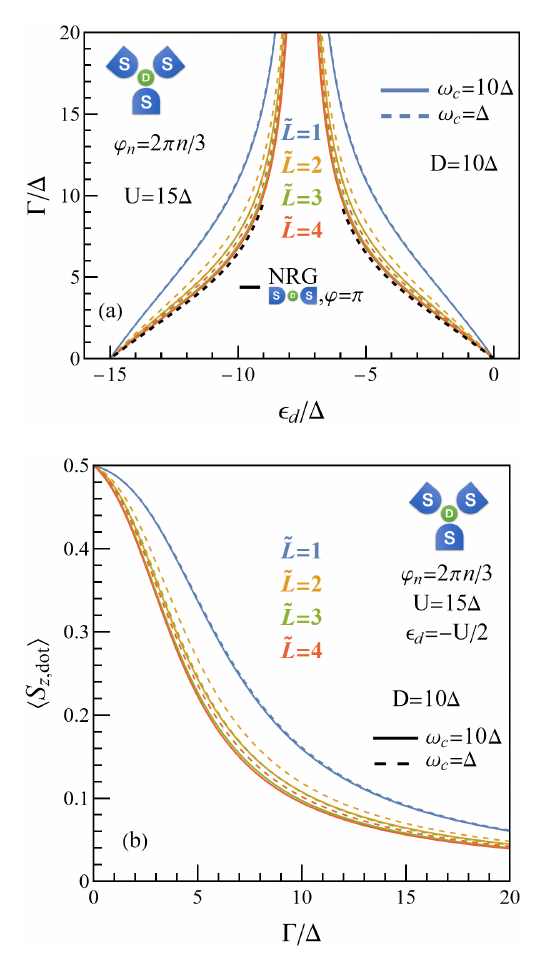}
\caption{Same as Fig.~\ref{figS10} but showing also the $\omega_c=\Delta$ results as dashed curves.}
\label{figS11}
\end{figure}

\begin{figure}
\includegraphics[width=\columnwidth]{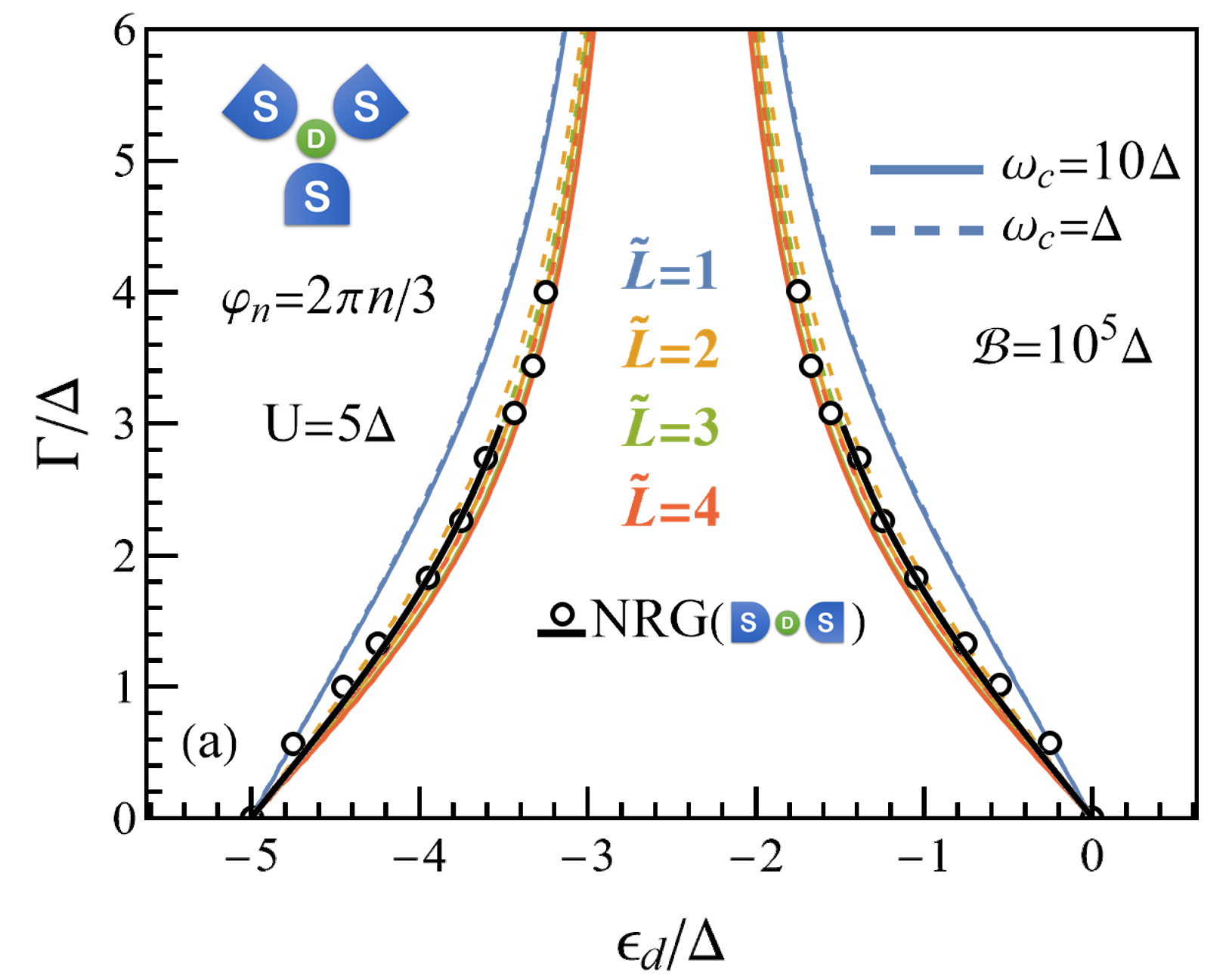}
\caption{Same as Fig.~\ref{figS10} but for a larger half bandwidth $\mathcal{B}=D=10^5\Delta$. The dots indicate the NRG data read graphically from Ref. \cite{Tanaka2007May} corresponding to the same bandwidth $D=10^5\Delta$, while the NRG black curve is obtained from Ref. \cite{rokdata} for $D=10\Delta$ and serves to evaluate the finite-bandwidth effects.}
\label{figS12}
\end{figure}

\clearpage

\begin{widetext}

\subsection{SMS with SC-SC direct coupling}

Considering a discretization in $L\gg 1$ levels of the SC leads, the total Hamiltonian for the 3-terminal device considered in the main text is given by
\begin{equation}
H=H_{\text{QD}}+\sum_{\beta=1}^3H_{\text{BCS},\beta}+\sum_{\beta=1}^3 \sum_{j=1}^L \sum_\sigma \left(t  \frac{1}{\sqrt{L}}c^\dagger_{\beta j\sigma}d_{\sigma} +h.c.\right)+\sum_{\beta=1}^3 \sum_{j,k=1}^L \sum_\sigma \left(r e^{i\alpha} \frac{1}{\sqrt{L}}\frac{1}{\sqrt{L}} c^\dagger_{\beta+1, j,\sigma}c_{\beta k\sigma} +\text{h.c.}\right)~,
\end{equation}
where $t$ ($r$) is the QD-SC (SC-SC) hopping amplitude. We build the SMS by sequentially replacing each SC lead by its corresponding surrogate, ending up with
\begin{equation}
\widetilde{H}=H_{\text{QD}}+\sum_{\beta=1}^3 H_{\widetilde{\text{SC}},\beta}+\sum_{\beta=1}^3 \sum_{\ell=1}^{\tilde L}\sum_{ \sigma=\uparrow \downarrow}\sqrt{\gamma_\ell \Gamma}( \tilde{c}_{\beta\ell \sigma}^{\dagger} d_{ \sigma}+\text {h.c.})+\sum_{\beta=1}^3 \sum_{\ell,\ell'=1}^{\tilde L}\sum_{ \sigma=\uparrow \downarrow}\Gamma_S\sqrt{\gamma_\ell \gamma_{\ell'}}( e^{i\alpha}\,\tilde{c}_{\beta+1,\ell, \sigma}^{\dagger} \tilde{c}_{\beta\ell' \sigma}+\text {h.c.})~,
\end{equation}
where $\Gamma=\pi t^2/2D$, $\Gamma_S=\pi r/2D$ and $\tilde{c}_{4,\ell,\sigma}\equiv \tilde{c}_{1,\ell,\sigma}$.

\begin{figure}[ht!]
\centering
\includegraphics[width=\columnwidth]{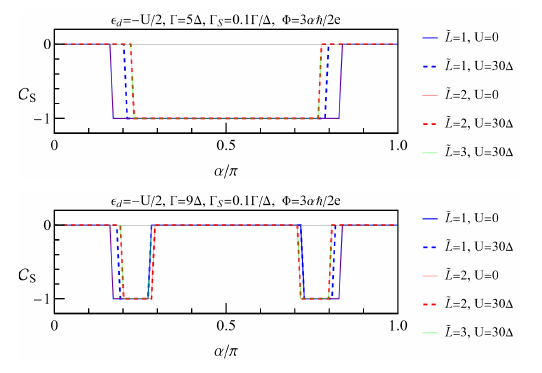}
\caption{Chern number convergence for the three-terminal device of Fig. 5 in the main text, with increasing number of effective levels $\tilde{L}$, in the non-interacting case ($U=0$) and for a finite (large) Coulomb interaction ($U=30\Delta$). The Chern number has been evaluated at $\alpha_n=(n-1)\pi/99$, $n=1,2,...,100$.}
\label{fig_ch}
\end{figure}

\begin{figure}[ht!]
\centering
\includegraphics[width=\columnwidth]{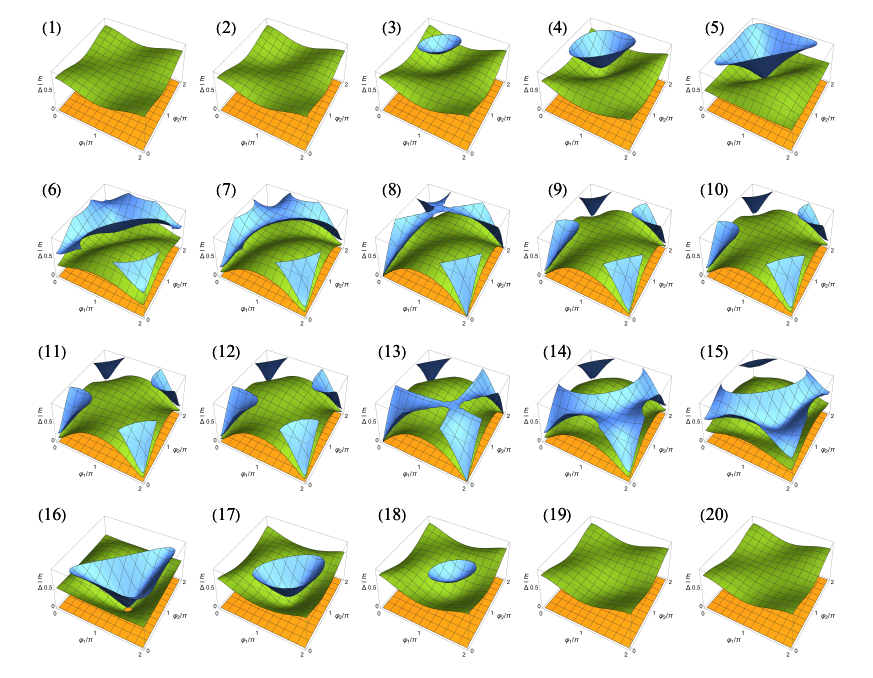}
\caption{Energy spectrum evolution with the enclosed magnetic flux $\alpha_n=(n-1)\pi/19$, $n=1,2,...,20$, for the three-terminal junction of Fig. 5 in the main text at the same parameters $\Gamma=7\Delta$, $U=30\Delta$, $\epsilon_d=-U/2$, $\Gamma_S=0.1\Gamma/\Delta$, $D=10\Delta$, $\tilde{L}=2$.  Orange indicates the lowest (reference) singlet, blue the first excited singled, green the lowest doublet. For the efficient scan of the $\varphi_1-\varphi_2$ first Brillouin zone we employed the reduced-basis method, see Refs. \cite{Baran2023Apr, Herbst2022Apr, Brehmer2023Apr} and discussion above. }
\label{fig20}
\end{figure}

\end{widetext}

\clearpage

\end{document}